\documentstyle[12pt,psfig,axodraw,a4]{article} 
\def\bild#1#2{    
        \vspace*{-5mm}
        \begin{center}
        \begin{math}
        \epsfxsize#2cm
        \epsffile{#1}
        \end{math}
        \end{center}
        }
\newcommand{\vs}{\vspace{-0.25cm}}

\begin{document}

\begin{center}
\Large{\bf Isovector nuclear spin-orbit interaction \\ from chiral pion-nucleon
dynamics}

\bigskip

N. Kaiser\\

\bigskip

{\small Physik Department T39, Technische Universit\"{a}t M\"{u}nchen, D-85747
Garching, Germany\\

\smallskip

{\it email: nkaiser@physik.tu-muenchen.de}}

\end{center}

\bigskip

\begin{abstract}
Using the two-loop approximation of chiral perturbation theory, we calculate 
the momentum and density dependent isovector nuclear spin-orbit strength 
$V_{ls}(p,k_f)$. This quantity is derived from the spin-dependent part of the 
interaction energy $\Sigma_{spin} = {i\over 2}\,\vec \sigma \cdot (\vec q 
\times\vec p\,)[U_{ls}(p,k_f)- V_{ls}(p,k_f)\tau_3 \delta] $ of a nucleon 
scattering off weakly inhomogeneous isospin-asymmetric nuclear matter. We find 
that iterated $1\pi$-exchange generates at saturation density, $k_{f0}=272.7\,
$MeV, an isovector nuclear spin-orbit strength at $p=0$ of $V_{ls}(0,k_{f0})
\simeq 50$\,MeVfm$^2$. This value is about 1.4 times the analogous isoscalar 
nuclear spin-orbit strength $U_{ls}(0,k_{f0})\simeq 35$\,MeVfm$^2$ generated by
the same two-pion exchange diagrams. We also calculate several 
relativistic $1/M$-corrections to the isoscalar nuclear spin-orbit strength. In
particular, we evaluate the contributions from irreducible two-pion exchange to
$U_{ls}(p,k_f)$. The effects of the three-body diagrams constructed from
the Weinberg-Tomozawa $\pi\pi NN$-contact vertex on the isoscalar nuclear 
spin-orbit strength are computed. We find that such relativistic 
$1/M$-corrections are less than 20\% of the isoscalar nuclear spin-orbit 
strength generated by iterated one-pion-exchange, in accordance with the
expectation from chiral power counting.
\end{abstract}


\vskip 0.5cm

\section{Introduction}
The introduction of the spin-orbit term into the nuclear single-particle
Hamiltonian has been most decisive for the success of the nuclear shell model 
\cite{haxel}. Only with a very strong and attractive spin-orbit potential one 
is, for example, able to explain the observed sequence of so-called magic 
numbers $\{2,8,20,28,50,82,126,\dots\}$. The dynamical origin of the strong 
nuclear spin-orbit force has not been fully resolved even up to date. The 
analogy with the spin-orbit interaction in atomic physics gave the hint that it
could be a relativistic effect. This idea has lead to the construction of the 
relativistic (scalar-vector) mean-field model \cite{walecka}. In this model the
spin-independent nuclear potential (of approximate depth $-50$\,MeV) results 
from an almost complete cancelation of a very strong attraction generated by 
scalar ($\sigma$-meson) exchange and a nearly equally strong repulsion 
generated by vector ($\omega$-meson) exchange. 
The corresponding spin-orbit term (obtained by a non-relativistic reduction of
the nucleon's Dirac-Hamiltonian) comes out proportional to the coherent sum of
the very large scalar and vector mean-fields. In this sense, the relativistic 
mean-field model gives a simple and natural explanation of the basic features 
of the nuclear shell model potential. Refinements of relativistic mean-field 
models which include additional non-linear couplings of the scalar and vector 
fields or explicitly density-dependent point couplings of nucleons are nowadays
widely and successfully used for nuclear structure calculations
\cite{ring,typel,lenske}.

The nuclear spin-orbit potential arises generally as a many-body effect from 
the underlying spin-orbit term in the (free) nucleon-nucleon scattering 
amplitude. The calculation of the tree level diagrams with one scalar-meson or
one vector-meson exchange between nucleons gives indeed a spin-orbit term in
the NN T-matrix proportional to $1/M^2$, with $M$ denoting the nucleon mass. 
The  nuclear spin-orbit potential corresponding to scalar and vector meson
exchange is therefore obviously a truly relativistic effect. However, the 
quadratic reciprocal scaling of the spin-orbit NN-amplitude with the nucleon 
mass $M$ is not universal, and it changes if one considers the exchange of two
mesons between nucleons, i.e. loop diagrams. For example, irreducible two-pion
exchange gives rise to a spin-orbit term in the NN T-matrix proportional to 
$1/M$ (see eqs.(22,23) in ref.\cite{nnpap1}) and iterated one-pion exchange
produces a spin-orbit term in the NN T-matrix which even scales linearly with
the nucleon mass $M$ (see eq.(33) in ref.\cite{nnpap1}). 

In a recent work \cite{spinbahn} we have used the systematic framework of
chiral perturbation theory to calculate the nuclear spin-orbit interaction
generated by one- and two-pion exchange. The momentum and density dependent 
nuclear spin-orbit strength $U_{ls}(p,k_f)$ is derived from the spin-dependent 
part of the interaction energy $\Sigma_{spin} = {i\over 2}\,\vec \sigma \cdot 
(\vec q \times\vec p\,)\,U_{ls}(p,k_f)$ of a nucleon scattering off weakly 
inhomogeneous isospin-symmetric nuclear matter. When working to linear order in
small spatial gradients (which corresponds to linear order in the momentum 
transfer $\vec q$\,) no detailed knowledge about the density distribution of 
nuclear matter is required. In fact, after isolating the important
proportionality factor $\vec q$ from the expressions of self-energy-type 
diagrams the nuclear spin-orbit strength $U_{ls}(p,k_f)$ can be finally
computed in the limit of homogeneous isospin-symmetric nuclear matter
(characterized by its Fermi momentum $k_f$). It was found in
ref.\cite{spinbahn} that the Hartree and Fock diagrams of iterated one-pion
exchange generate a spin-orbit strength at $p=0$ of $U_{ls}(0,k_{f0})=34.7 
$\,MeVfm$^2$ which is in perfect agreement with the empirical value $\widetilde
U_{ls}\simeq 35\,$MeVfm$^2$ \cite{bohr,eder} used in shell-model calculation 
of nuclei. This novel spin-orbit strength is neither of relativistic nor of 
short range origin. It is in fact linearly proportional to the nucleon mass 
$M$ and its inherent range is the pion Compton wavelength $m_\pi^{-1}=1.46 \,
$fm. A strong $p$-dependence of the nuclear spin-orbit strength $U_{ls}(p,
k_{f0})$ arising from iterated $1\pi$-exchange has however been observed in 
ref.\cite{spinbahn}. A further property of the spin-orbit Hamiltonian emerging 
from that diagrammatic calculation was that it has (in coordinate space) terms 
proportional to $\vec \nabla f(r)$ as well as terms proportional to $f(r)\,\vec
\nabla f(r)$ (with $f(r)=\rho(r)/\rho(0)$ the normalized radial density 
profile) which get weighted differently at the surface of a finite nucleus. All
such (unusual) features which go beyond the simple shell model parameterization
of the nuclear spin-orbit Hamiltonian leave questions about the ultimate 
relevance of the spin-orbit interaction generated by $2\pi$-exchange for a
finite nucleus. Nuclear structure calculation which use the results of
ref.\cite{spinbahn} as input are necessary in order to clarify the role of the
nuclear spin-orbit interaction generated by $2\pi$-exchange. 

The purpose of the present work is to calculate using the same framework as in 
ref.\cite{spinbahn} the isovector spin-orbit strength $V_{ls}(p,k_f)$ generated
by iterated $1\pi$-exchange in order to reveal the underlying isospin 
dependence. Furthermore, we calculate here several relativistic 
$1/M$-corrections to the isoscalar nuclear spin-orbit strength $U_{ls}(p,k_f)$.
In particular, we evaluate the contributions to $U_{ls}(p,k_f)$ arising from 
irreducible two-pion exchange. We compute the isoscalar nuclear spin-orbit 
strength generated by certain three-body diagrams constructed from the leading 
order isovector chiral $\pi\pi NN$-contact vertex (so-called Weinberg-Tomozawa
vertex). The relativistic $1/M$-corrections to the dominant Hartree diagrams of
iterated $1\pi$-exchange are also calculated. These results shed some light on
the convergence of the small momentum expansion underlying the whole approach. 

\section{General considerations about the nuclear spin-orbit term}
Let us begin with recalling the spin-orbit Hamiltonian of the nuclear shell 
model \cite{bohr,eder} which is generally written in the form:
\begin{equation} {\cal H}_{ls}= \Big(\widetilde U_{ls}- \widetilde V_{ls}\,
\tau_3 \delta \Big)\,{\vec \sigma \cdot\vec \ell
\over 2 r }\, {d f(r) \over d r}\,.  \end{equation}
Here, $ \vec \sigma$ is the conventional Pauli spin-vector and $\vec \ell = 
\vec r \times \vec p $ is the orbital angular momentum of a nucleon. $f(r)$ 
denotes the normalized radial density distribution of a nucleus, typically 
parameterized by a Saxon-Woods function. $\delta =(N-Z)/(N+Z)$ is the isospin
asymmetry parameter and $\tau_3 \to \pm1$ for a proton or a neutron. The
empirical value of the isoscalar nuclear spin-orbit strength is  $\widetilde
U_{ls} \simeq 35\,{\rm MeV fm}^2$ \cite{bohr,eder}. The ratio of the isovector
and isoscalar nuclear spin-orbit strengths $\widetilde V_{ls}/\widetilde U_{ls}
\simeq -0.6$ is commonly assumed to be equal to the ratio of the 
spin-independent isovector and isoscalar average nuclear potentials 
\cite{bohr,eder}. 
 
We wish to calculate the nuclear spin-orbit strengths $\widetilde U_{ls}$ and
$\widetilde V_{ls}$ (or appropriate generalization of them) in the systematic 
framework of in-medium chiral perturbation theory \cite{einpot}. The first 
observation one makes is that the spin-orbit interaction vanishes identically 
in infinite homogeneous nuclear matter since there is no preferred center in 
this system in order to define an orbital angular momentum. Therefore one has 
to generalize the calculation of the single-particle potential in 
ref.\cite{einpot} to (at least) weakly inhomogeneous nuclear matter. The 
relevant quantity in order to extract the isoscalar and isovector nuclear 
spin-orbit strengths is the spin-dependent part of the interaction energy of a
nucleon scattering off weakly inhomogeneous isospin-asymmetric nuclear matter
from initial momentum $\vec p-\vec q/2$ to final momentum $\vec p+\vec q/2$,
which  reads: 
\begin{equation}  \Sigma_{spin} = {i\over 2}\, \vec \sigma \cdot (\vec q \times
\vec p\,) \,\Big\{ U_{ls}(p,k_f)-  V_{ls}(p,k_f)\, \tau_3 \,\delta +{\cal
O}(\delta^2)\Big\}\,. \end{equation} 
The (small) momentum transfer $\vec q$ is provided by the Fourier-components 
of the inhomogeneous nuclear matter distribution. The density form factor 
$\Phi(\vec q\,)= \int d^3 r\, e^{-i \vec q \cdot \vec r} f(r)$ plays the
role of a probability distribution of these Fourier-components. With the help
of this relationship the spin-orbit Hamiltonian ${\cal H}_{ls}$ in eq.(1) 
becomes equal to the Fourier-transform of the product of the density form 
factor $\Phi(\vec q\,)$ and the spin-dependent interaction energy $\Sigma_{
spin}$: ${\cal H}_{ls} =(2\pi)^{-3}\int d^3 q \,e^{i \vec q \cdot \vec r}\,
\Phi(\vec q\,)\,\Sigma_{spin}$. 

\bigskip
\begin{center}
\bild{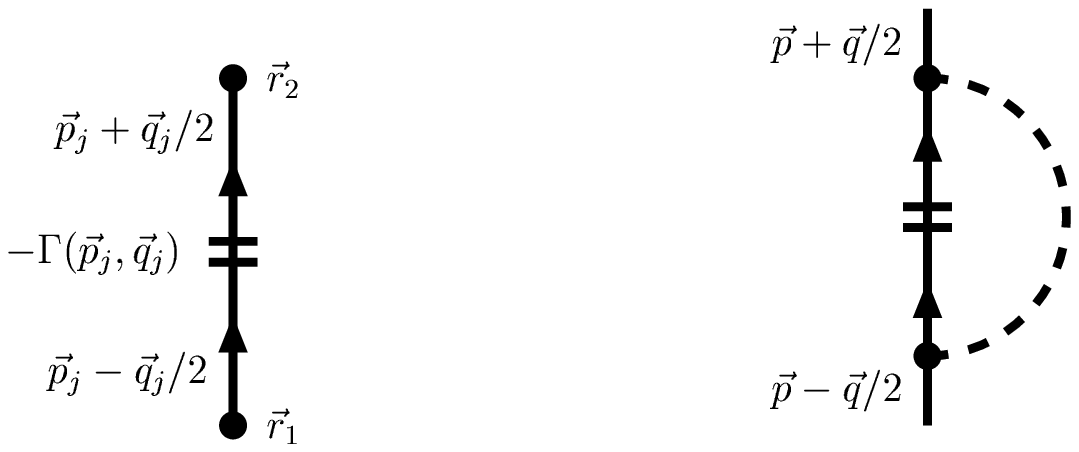}{18.7}
\end{center}
{\it Fig.1: The double line symbolizes the medium insertion for a weakly  
inhomogeneous many-nucleon system $\Gamma(\vec p_j,\vec q_j)$ defined by
eqs.(3,4). Right: The $1\pi$-exchange Fock graph.}

\bigskip
Consistent with the assumption of a weakly
inhomogeneous nuclear matter distribution we keep in $\Sigma_{spin}$ only
linear terms in $\vec q$ corresponding to small spatial gradients of the
density profile $f(r)$. Practically, this means that after isolating the
proportionality factor $\vec q$ in an explicit calculation one can take the
limit of homogeneous isospin-asymmetric nuclear matter characterized by its
proton and neutron Fermi momenta $k_{p,n}=k_f(1\mp \delta)^{1/3}$. The
expansion up to linear order in the isospin asymmetry parameter $\delta$ gives
then the momentum and density dependent isoscalar and isovector nuclear
spin-orbit strengths $ U_{ls}(p,k_f)$ and $V_{ls}(p,k_f)$. Possible higher
powers in $\vec q$ contributing to $\Sigma_{spin}$ have no direct
correspondence to the (standard) spin-orbit Hamiltonian ${\cal H}_{ls}$ in
eq.(1).     

In ref.\cite{einpot} the calculation of the spin-independent single-particle 
potential in homogeneous isospin-symmetric nuclear matter has been 
organized in the number of so-called medium insertions. The medium insertion
for a (non-relativistic) many-fermion system is generally constructed from the
sum over the occupied  energy eigenstates as \cite{gross,vauth}:
\begin{equation} \Gamma(\vec p_j,\vec q_j) = \int d^3r_1 \int d^3r_2\sum_{E\in 
occ} \psi_E(\vec r_2) \psi^*_E(\vec r_1) \, e^{i \vec p_j \cdot (\vec r_1
-\vec r_2)} \,  e^{-{i\over 2} \vec q_j \cdot (\vec r_1+\vec r_2)} \,.
\end{equation} 
The double line in the left picture of Fig.\,1 symbolized this medium insertion
together with the assignment of the out- and in-going nucleon momenta $\vec p_j
\pm\vec q_j/2$. The momentum transfer $\vec q_j$ is provided by the 
Fourier-components of the inhomogeneous matter distribution. Semi-classical
expansions \cite{gross,vauth} give for a weakly inhomogeneous and
spin-saturated many-nucleon system: 
\begin{equation} \Gamma(\vec p_j,\vec q_j) = \bigg[{1+\tau_3 \over 2}\,\theta(
k_p - |\vec p_j|)+ {1-\tau_3 \over 2}\,\theta(k_n - |\vec p_j|)\bigg] 
\Phi(\vec q_j ) \Big\{ 1+ {\cal O}(\vec q_j )\Big\} \,,  \end{equation}  
with $\Phi(\vec q_j)$ the density form factor introduced after eq.(2). The 
subleading ${\cal O}(\vec q_j)$ term in eq.(4) will never come into play in the
calculation of the spin-dependent interaction energy $\Sigma_{spin}$ to linear
order in $\vec q$.   

\section{Diagrammatic calculation of the isovector nuclear \\
spin-orbit strength}

\bigskip
\bild{isorbitfig1.epsi}{14}

{\it Fig.2: The iterated $1\pi$-exchange Hartree graphs. These diagrams are  
labeled (a), (b), (c), (d) from left to right.}

\bigskip

In this section we present analytical results for the isovector nuclear 
spin-orbit strength $V_{ls}(p,k_f)$ as given by chiral one- and two-pion 
exchange. We start with the $1\pi$-exchange Fock\footnote{Our nomenclature of
"Hartree" and "Fock" diagrams is taken over from ref.\cite{einpot}. There
closed $1\pi$- and $2\pi$-exchange vacuum diagrams with two (resp. one) closed
nucleon line have been called Hartree (resp. Fock) diagrams. The one-particle
property $\Sigma_{spin}$ is obtained by opening one nucleon line of such
diagrams. Consequently, we call here diagrams with one open and one closed 
nucleon line "Hartree" diagrams and diagrams with only one open nucleon line
"Fock" diagrams.} graph (right diagram in
Fig.\,1). In the static approximation the product of $\pi N$-interaction
vertices $\vec \sigma \cdot (\vec p - \vec p_1)\,  \vec \sigma \cdot (\vec p -
\vec p_1)= (\vec p -\vec p_1)^2$ is spin-independent. A non-vanishing nuclear 
spin-orbit strength comes therefore only as a relativistic $1/M^2$-correction. 
Isolating the $ i \, \vec \sigma \times \vec q$ factor from the product of 
fully relativistic  pseudo-vector $\pi N$-interaction vertices, performing the 
$1/M$-expansion, integrating over Fermi-spheres of radii $k_{p,n} = k_f(1\mp
\delta)^{1/3}$ and expanding to linear order in the asymmetry parameter
$\delta$, we get from the $1\pi$-exchange Fock diagram: 
\begin{equation} V^{(1\pi)}_{ls}(p,k_f)= {2g_A^2m_\pi^3 u^2(u^2-x^2) \over 3
(8\pi f_\pi M x)^2 }\Big\{(1+u^2-x^2) L(x,u)- u\Big\}  \,. \end{equation} 
Here, we have introduced the dimensionless variables $x=p/m_\pi$ and
$u=k_f/m_\pi$ and the auxiliary function 
\begin{equation}L(x,u)={1\over4x}\ln{1+(u+x)^2\over 1+(u-x)^2}\,.\end{equation}

Next, we come to the evaluation of the four Hartree diagrams of iterated 
one-pion exchange shown in Fig.\,2. We start with the left graph with one
medium insertion, labeled (a). The  relevant $i\,\vec \sigma \times \vec q$ 
prefactor can be isolated already in the first step of the calculation from the
product of $\pi N$-interaction vertices $\vec\sigma\cdot(\vec l-\vec q/2)\,\vec
\sigma \cdot (\vec l + \vec q/2)$ at the open nucleon line. For all remaining
parts of that diagram one can then take the limit of homogeneous
isospin-asymmetric nuclear matter (i.e. $\vec q=0$). After solving the inner
$d^3l$-loop integral \cite{nnpap1} and expanding the integrals over Fermi 
spheres of radii $k_{p,n} = k_f(1\mp \delta)^{1/3}$ to linear order in
$\delta$, we find the following closed form expression for the isovector
nuclear spin-orbit strength generated by the Hartree diagram (a):   
\begin{equation} V^{(a)}_{ls}(p,k_f)= {g_A^4 M m_\pi^2 u^2\over (4\pi x)^3 
f_\pi^4}\bigg\{ {8u\over 3}\Big[\arctan(u-x) -\arctan(u+x) \Big] +2x \bigg( 
{5\over 3}-u^2+x^2 \bigg) L(x,u) +2ux \bigg\} \,.  \end{equation}
We continue with the calculation of the Hartree diagrams with two medium
insertions, labeled (b), (c) and (d) in Fig.\,2. In these diagrams, a (small)
momentum transfer $\vec q_{1,2}$ with $\vec q=\vec q_1+\vec q_2$ occurs at each
medium insertion. The important prefactors $i\,\vec \sigma \times \vec q_j$ can
again be isolated in the first step of the calculation from the product of $\pi
N$-interaction vertices at the open nucleon line. After that the vectors $\vec
q_j$ can be set to zero in all remaining components of these diagrams. When 
Fourier-transformed with density form factors to coordinate space each such
momentum transfer $\vec q_j,\,(j =1,2)$ leads to the expression: $(2\pi)^{-6}
\int d^3 q \int d^3 q_j \,e^{i \vec q\cdot \vec r} \, i\vec q_j \,\Phi(\vec
q_j)\,\Phi(\vec q - \vec q_j) = f(r) \, \vec \nabla f(r)$, with $f(r)$ 
the density profile of weakly inhomogeneous nuclear matter. Consistent with the
assumption of a weakly inhomogeneous nuclear matter (i.e. keeping only linear 
terms in small spatial gradients) we can make the replacement: $f(r) \, \vec 
\nabla f(r) \to \vec \nabla f(r)$. The essential conclusion from this 
consideration is that for the calculation of $\Sigma_{spin}$ in weakly 
inhomogeneous nuclear matter each momentum transfer $\vec q_j$ can be 
identified with $\vec q$. With the help of this rule and certain techniques to 
reduce six-dimensional principal value integrals over the product of two 
Fermi-spheres, we end up after expanding to linear order in $\delta$ with the 
following result for the  Hartree diagram (b):
\begin{eqnarray} V^{(b)}_{ls}(p,k_f)&=& {g_A^4 M m_\pi^2\over (4\pi f_\pi)^4}
\int_{-1}^1 dy \,{4y\over 3x} \bigg\{2u^2 \bigg[ {3s+2s^3 \over 1+s^2} -3 
\arctan s \bigg] \ln{u+x y\over u-x y}\nonumber \\ && -{ s' s^4 \over (1+s^2)^2
} \bigg[2u x y +(u^2-x^2y^2)\ln{u+x y\over u-x y} \bigg] \bigg\}\end{eqnarray} 
with the auxiliary functions $s=x y +\sqrt{u^2-x^2+x^2y^2}$ and $s' = u\,
\partial s/\partial u$. Throughout this work the momentum $p$ is restricted to
the interval $0 \leq p \leq k_f$. Similarly, we find for the Hartree diagram  
(c):
\begin{eqnarray} V^{(c)}_{ls}(p,k_f)&=& {g_A^4 M m_\pi^2\over (4\pi f_\pi)^4}
\int_{-1}^1 dy \, \int_{-x y}^{s-xy} d\xi\,{4y(xy+\xi)^4\over 3s x[1+(xy+\xi)^2
]^2}\, \bigg\{s'(3\xi+ 2xy)\bigg[\xi \ln{u+\xi\over u-\xi}-2u \bigg]\nonumber 
\\ && - {4s' \over 1+(xy+\xi)^2} \bigg[2u \xi +(u^2-\xi^2) \ln{u+\xi\over u-
\xi} \bigg]-(4s+s') u^2\ln{u+\xi\over u-\xi} \bigg\}\,,\end{eqnarray} 
and for the Hartree diagram (d):
\begin{eqnarray} V^{(d)}_{ls}(p,k_f)&=& {g_A^4 M m_\pi^2\over (2\pi f_\pi)^4}
\int_0^u d\xi\,{2\xi^2 \over 3x^3} -\!\!\!\!\!\!\int_{-1}^1 dy \bigg\{ {x^3
\over x^2-\xi^2 y^2}\bigg[ {3\sigma+2\sigma^3 \over 1+\sigma^2} -3 \arctan 
\sigma \bigg] \nonumber\\ && + \bigg[\xi y \ln{|x+\xi y|\over |x-\xi y|} -2x 
\bigg] \bigg[ {\sigma(1-2 \sigma^3 \sigma') \over (1+\sigma^2)^2} +5\sigma - 
6\arctan\sigma\bigg] \bigg\}\,, \end{eqnarray} 
with the auxiliary functions $\sigma =\xi y +\sqrt{u^2-\xi^2+\xi^2y^2}$ and
$\sigma' = u \, \partial \sigma/\partial u$. The symbol $-\!\!\!\!\!\int_{-1}^1
dy$ in eq.(10) stands for a principal value integral. Note the similarity of 
the expressions eqs.(8,9,10) with the three-body isovector single-particle 
potentials written in eqs.(30,31,32) of ref.\cite{spinbahn}. 

\bigskip

\bild{isorbitfig2.epsi}{14}
{\it Fig.3: The iterated $1\pi$-exchange Fock graphs. These diagrams are 
labeled (e), (f), (g), (g) from left to right.}

\bigskip

Next, we come to the evaluation of the four iterated $1\pi$-exchange Fock 
diagrams shown in Fig.\,3. We start with diagram (e) with one medium insertion.
The method to isolate the important proportionality factor $\vec q$ has been
explained in section 3 of ref.\cite{spinbahn}. Considering the isospin 
factor of diagram (e) and expanding the integrals over Fermi spheres of radii
$k_{p,n} = k_f(1\mp \delta)^{1/3}$ to linear order in the asymmetry parameter
$\delta$, we end up with the following contribution to the isovector nuclear 
spin-orbit strength:
\begin{eqnarray} V^{(e)}_{ls}(p,k_f)&=& {5g_A^4 M m_\pi^2 u^2\over 3(8\pi x)^3 
f_\pi^4}\bigg\{2(u^2-x^2) \ln{u+x\over u-x}-4u x + \int_{(u-x)/2}^{(u+x)/2}
d\xi \,{ 4\xi^2+x^2-u^2\over \xi^2 (1+2\xi^2)}\nonumber \\ && \times \Big[(1+
4\xi^2) \arctan 2\xi -4\xi^2(1+\xi^2) \arctan \xi \Big] \bigg\} \,.
\end{eqnarray} 
We continue with the computation of the Fock diagram (f) with a symmetrical 
arrangement of the two medium insertions. It leads to the following 
contribution to the isovector nuclear spin-orbit strength: 
\begin{eqnarray} V^{(f)}_{ls}(p,k_f)&=& {g_A^4 M m_\pi^2\over (4\pi f_\pi)^4}
\bigg\{ {u^2 \over 3x^4} G(x,u) \Big[(u^2+1-2x^2) L(x,u)-u \Big]\nonumber \\ &&
+ {u \over 12x^4} {\partial G(x,u) \over \partial u} \bigg[2x^2\Big(\arctan
(u+x)+\arctan(u-x)\Big)\nonumber \\ && -u(1+u^2+3x^2) +\Big((1+u^2)^2+x^2(3x^2
-4u^2)\Big) L(x,u) \bigg]  \nonumber \\ && +\int_{-1}^1 dy\, \int_{-1}^1 dz \,
{yz \,\theta(y^2+z^2-1) \over 3x|yz| \sqrt{y^2+z^2-1}} \,{s^2 s'\over 1+s^2}
 \nonumber \\ && \times \Big[2sz( t-\arctan t) +y(t^2-\ln(1+t^2) \Big]\bigg\}
 \,, \end{eqnarray}  
with the auxiliary functions $G(x,u) =u(1+u^2+x^2)-[1+(u+x)^2] [1+(u-x)^2] 
L(x,u)$ and  $t=x z +\sqrt{u^2-x^2+x^2z^2}$. Finally, we have to evaluate the 
last two topologically distinct Fock diagrams in Fig.\,3. Since they
contribute equally to the spin-dependent interaction energy $\Sigma_{spin}$ 
we have given both diagrams the same label (g). We find the following 
representation for their contribution to the isovector nuclear spin-orbit
strength: 
\begin{equation} V_{ls}^{(g)}(p,k_f) = -{g_A^4 M m_\pi^2 u\over 6x^3 (4\pi 
f_\pi)^4}\bigg[ H(x,u,u)+5 \int_0^u d\xi \,{\partial H(x,u,\xi) \over \partial
u} \bigg]\,. \end{equation}    
Here, we have introduced the abbreviation: 
\begin{eqnarray} H(x,u,\xi)&=& 3ux \xi^{-2}(1+u^2)(1+\xi^2)-ux (12\xi^2+1+x^2) 
 \nonumber \\ && +8\xi^2\Big[\arctan(u+\xi)+\arctan(u-\xi) 
\Big]\Big[x+(x^2-1-\xi^2)L(\xi,x) \Big]  \nonumber \\ &&  +\Big[ 3\xi^4 
(4u^2+5x^2-3) +\xi^2 (5+10u^2-4x^2-14u^2x^2-x^4)-12\xi^6-2u^2x^4
\nonumber \\ && +5+3u^4+3x^2-u^4x^2+4u^2+2x^4-6u^2x^2+3\xi^{-2}
(1+u^2)^2(1+x^2)^2 \Big] \nonumber \\ && \times L(\xi,x) L(\xi,u) +  \Big[   
12 \xi^4 +\xi^2 (x^2-12u^2-3)+ 2(1+x^2)(u^2-1)\nonumber \\ &&
-3 \xi^{-2} (1+x^2)(1+u^2)^2 \Big] x L(\xi,u) + \Big[ 12\xi^4 +\xi^2 
( 13-15x^2)\nonumber \\ && +x^4+u^2x^2 +3x^2-3u^2-2 
 -3 \xi^{-2}(1+x^2)^2(1+u^2) \Big]u L(\xi,x) \nonumber \\ && +\xi^2 \int_{-1}^1
 dy \bigg\{ 16(\sigma - \arctan\sigma)[x-L(\xi,x)]  \nonumber \\ &&  +\bigg[  
\ln(1+ \sigma^2) +8\xi y \arctan\sigma +4u^2-4\xi^2 -5 \sigma^2
\bigg] \ln{|x+\xi y|\over |x-\xi y|}\nonumber \\ &&  + \bigg[ 8\xi y \arctan
\sigma+ (1+\xi^2-x^2) \ln(1+ \sigma^2)  +4u^2-4\xi^2  \nonumber \\
&& +\sigma^2 (x^2-5-\xi^2) \bigg]  {1\over R } \ln{ |x R+
(x^2-1-\xi^2)\xi y| \over |x R+ (1+\xi^2-x^2)\xi y|} \bigg\}\,, \end{eqnarray}
for the combined integrand of eqs.(21,23) in ref.\cite{spinbahn} as well as the
auxiliary function $R=\sqrt{(1+x^2-\xi^2)^2+4\xi^2(1-y^2)}$. This completes the
presentation of analytical results for the isovector nuclear spin-orbit 
strength $V_{ls}(p,k_f)$ generated by $1\pi$-exchange and iterated 
$1\pi$-exchange.

\subsection{Results}
For the numerical evaluation of the isovector nuclear spin-orbit strength 
$V_{ls}(p,k_f)$ we use consistently the same parameters as in our previous 
works \cite{spinbahn,einpot}. We choose the value $g_A=1.3$ for the nucleon 
axial  vector coupling constant. The weak pion decay constant has the value 
$f_\pi =  92.4\,$MeV and $M=939\,$MeV and $m_\pi= 135\,$MeV are the masses of
the nucleon and the (neutral) pion, respectively.

\bigskip
\begin{table}
\begin{center}
\begin{tabular}{|c|cccccccc|}
    \hline

    Diagram & $1\pi$-Fock & (a) & (b) & (c) & (d) & (e)  & (f) & (g)  
     \\  \hline     $V_{ls}(0,k_{f0})$ 
& $-$0.14 &24.72 & $-$1.49 & $-$18.94 & 51.40 & $-$42.35 & $-$4.55 & 41.94  
\\    \hline
$V_{ls}(k_{f0},k_{f0})$
& 0.00 &22.03 & 10.60 & $-$10.64 & 25.83 &$-$38.40 & $-$2.68 & 27.04 
 \\    \hline
  \end{tabular}

\end{center}

{\it Table\,1: Contributions of individual diagrams to the isovector nuclear 
spin-orbit strength $V_{ls}(p,k_{f0})$ at $p=0$ and at $p=k_{f0}=272.7\,$MeV. 
The units are MeVfm$^2$.}
\end{table}

In the second row of Table\,1, we present numerical values for the
contributions of individual diagrams to the isovector nuclear spin-orbit 
strength $V_{ls}(0,k_{f0})$ at the nuclear matter saturation density $k_{f0}=
272.7\,$MeV (as predicted by our calculation \cite{einpot}). As expected 
the relativistic $1/M^2$-correction from the $1\pi
$-exchange Fock graph is a very small effect. The contributions of individual 
iterated $1\pi$-exchange diagrams are large and comparable in magnitude to the 
respective contributions to the isoscalar nuclear spin-orbit strength $U_{ls}(
0,k_{f0})$ (see Table\,1 in ref.\cite{spinbahn}). The basic reason for these 
large values is the large scale enhancement factor $M$ (the nucleon mass) 
entering the iterated $1\pi$-exchange. It stems from the energy denominator of 
such second-order diagrams which is a difference of small nucleon kinetic 
energies. Adding up the entries in the second row of Table\,1 one gets $V_{ls}
(0,k_{f0}) =50.6\,{\rm MeVfm}^2$, which is about 1.4 times the isoscalar 
nuclear spin-orbit strength $U(0,k_{k0})= 35.1\,{\rm MeVfm}^2$ \cite{spinbahn}
generated by the same $2\pi$-exchange diagrams. The predicted total sum is 
dominated by the contribution $V^{(H)}_{ls}(0,k_{f0}) =55.7\,{\rm MeVfm}^2$ of
the four Hartree diagrams (a), (b), (c) and (d) (see Fig.\,2). The same
feature, namely the numerical suppression of the iterated  $1\pi$-exchange Fock
diagrams against the Hartree diagrams, holds also for the isoscalar spin-orbit
strength $U_{ls}(0,k_{f0})$ \cite{spinbahn} as well as for the
(spin-independent) nuclear mean-field studied in ref.\cite{einpot}. The
numerical entries (a)-(g) in Table\,1 are all of the same order in the small
momentum expansion. Individual cancelations give therefore no hint on the
"convergence" of the chiral expansion of $V_{ls}(p,k_f)$. Note that
the isovector nuclear spin-orbit strength $V_{ls}(0,k_{f0})$ generated (almost
completely) by iterated  $1\pi$-exchange is neither of relativistic nor of
short range origin. It is in fact linearly proportional to the nucleon mass $M$
and its inherent range is the pion Compton wavelength $m_\pi^{-1}=1.46\,$fm. 
The full pion Compton wavelength $m_\pi^{-1}$ and not half of it is suggested 
by comparing the expression $V_{ls}^{(a)}(p,k_f)$ in eq.(7) with the
$1\pi$-exchange single-particle potential $U_2(p,k_f)$ in eq.(8) of
ref.\cite{einpot}. As a matter of fact the same non-polynomial functions 
$L(x,u)$ and $\arctan(u \pm x)$ occur in both cases. 

In Fig.\,4, we show by the full line the dependence of the calculated isovector
nuclear spin-orbit strength $V_{ls}(0,k_f)$ on the nucleon density $\rho=2k_f^3
/3\pi^2$. One observes in the region $\rho \leq 0.4\,$fm$^{-3}$ an approximate 
linear growth of $V_{ls}(0,k_f)$ with the density $\rho$ as it is typical for
relativistic mean-field models. In a finite nucleus the spin-orbit force acts 
mainly on the surface where the density gradients are largest and the density 
has dropped to about half of the central density. The replacement $f(r) \,\vec 
\nabla f(r)\to {1\over 2}\vec \nabla f(r)$ (instead of $f(r) \,\vec \nabla f(r)
\to \vec \nabla f(r)$ valid for weakly inhomogeneous nuclear matter) describes 
then more realistically the situation for a finite nucleus. The dashed line in 
Fig.\,4 shows the isovector nuclear spin-orbit strength which results if the 
contributions from the diagrams with two medium insertion (b), (c), (d), (f) 
and (g) are weighted with a factor $1/2$. This different weighting reduces of 
the total isovector spin-orbit strength by about a factor 3.

\bigskip

\bild{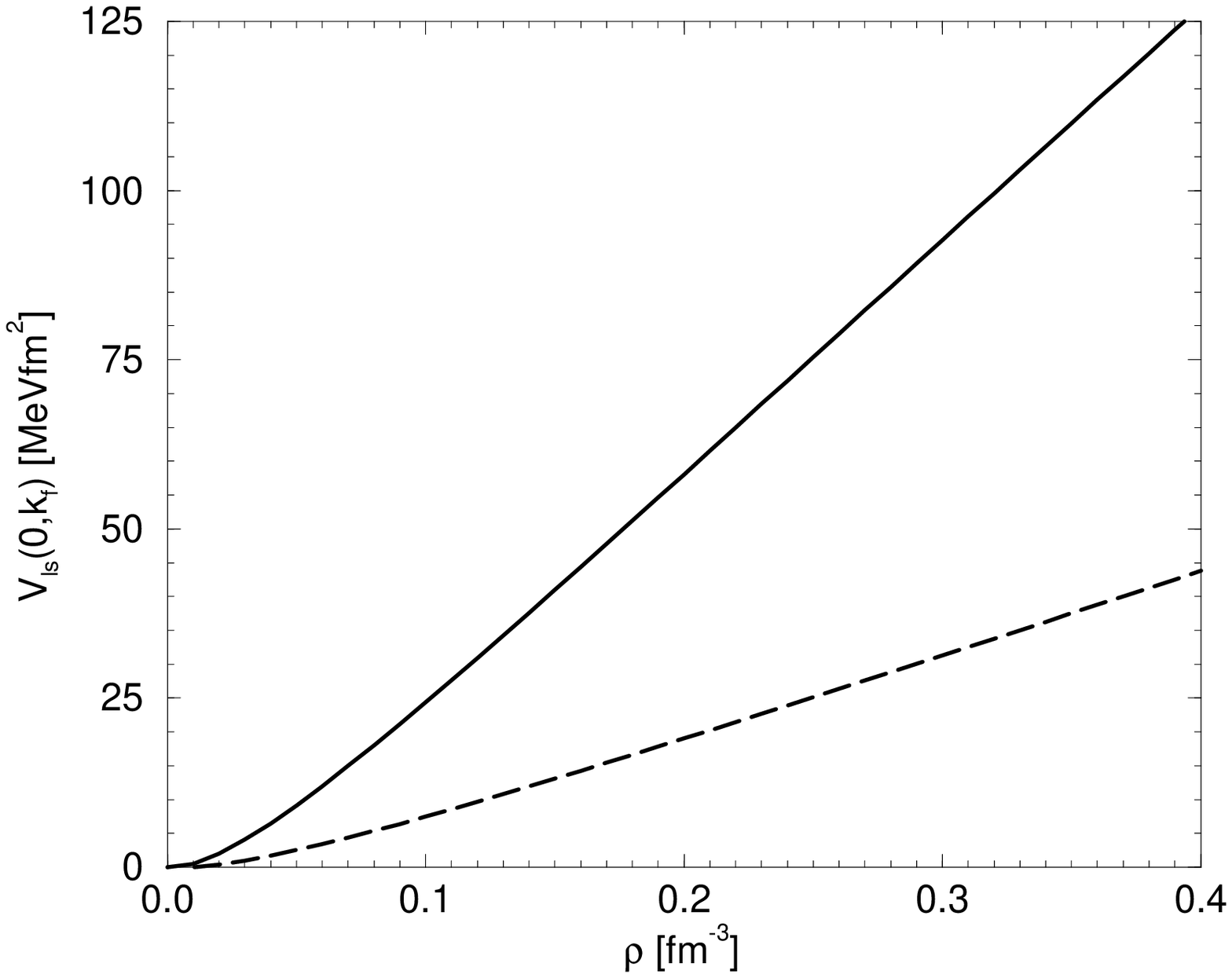}{13.}
\vspace{-1.2cm}
{\it Fig.\,4: The full line shows the isovector nuclear spin-orbit strength 
$V_{ls}(0,k_f)$ at zero nucleon momentum $(p=0)$ versus the nucleon density
$\rho= 2k_f^3/3\pi^2$. If the diagrams with two medium insertions are weighted
with a factor 1/2 the dashed line results.}

\bild{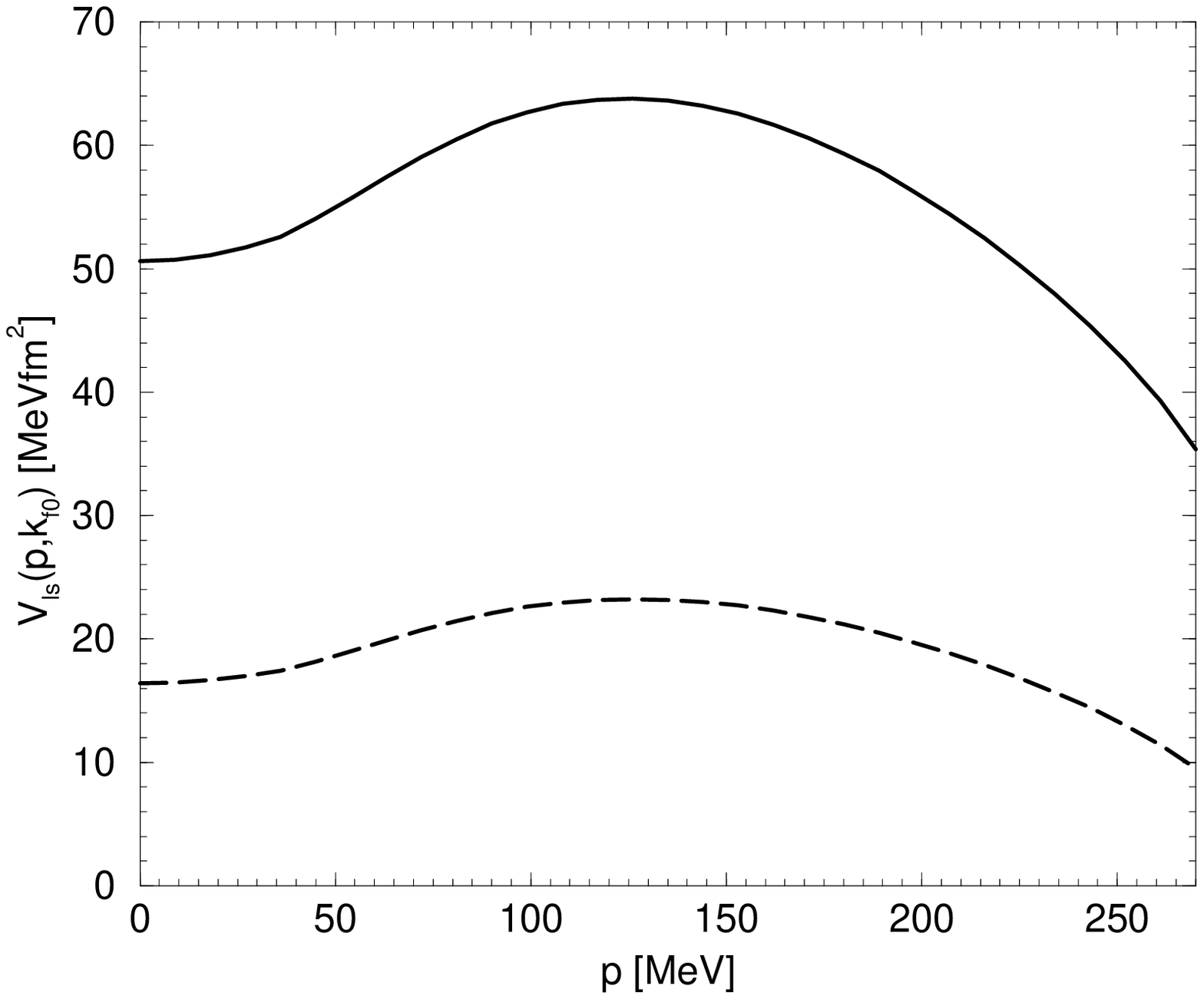}{13.}
\vspace{-1.2cm}
{\it Fig.\,5: The momentum dependence of the isovector nuclear spin-orbit
strength $V_{ls}(p,k_{f0})$ at saturation density $k_{f0}=272.7$\,MeV. The full
(dashed) line corresponds to a weighting of the diagrams with two medium
insertions with a factor 1 (1/2).}

\bigskip

In Fig.\,5, we show by the full line the dependence of the calculated isovector
nuclear spin-orbit strength $V_{ls}(p,k_{f0})$ at saturation density $k_{f0}=
272.7 $\,MeV on the nucleon momentum $p$ for $0\leq p\leq k_{f0}$. One 
observes a moderate $p$-dependence. In contrast to the isoscalar nuclear
spin-orbit strength $U_{ls}(p,k_{f0})$ shown in Fig.\,7 of ref.\cite{spinbahn}
there is no sign change in the present case. The dashed line in Fig.\,5 
corresponds to the weighting of diagrams with two medium insertions with a 
factor 1/2. It shows the same qualitative $p$-dependence as the full line in
Fig.\,7. The entries in the third row of Table\,1 indicate how the 
contributions from individual diagrams vary with the nucleon momentum from 
$p=0$ to $p=k_{f0}$. 

The terms from iterated $1\pi$-exchange written down in eqs.(7--14) are the
unique leading order contributions in the small momentum expansion. They are of
second order in small momenta as signaled by the common prefactor $g_A^4 M 
m_\pi^2/(2\pi f_\pi)^4$. Nuclear structure calculations which use the here
calculated isovector nuclear spin-orbit strength $V_{ls}(p,k_f)$ as input would
be most useful in order to clarify the role of the spin-orbit interaction
generated by $2\pi$-exchange in a finite nucleus.  
\section{Relativistic corrections to the isoscalar nuclear spin-orbit strength}
In this section we calculate several higher order corrections to the isoscalar
nuclear spin-orbit strength $U_{ls}(p,k_f)$ generated by $2\pi$-exchange in
order to learn about the convergence of the underlying small momentum 
expansion. All these contributions scale like $1/M$ with nucleon mass $M$ and
can therefore be regarded as truly relativistic spin-orbit effects. Since our
calculation is a strictly perturbative one $\pi N$-interaction vertices 
in-medium are the same as in vacuum. The relativistic spin-orbit effects depend
therefore reciprocally on the free nucleon mass $M$ and not on a
self-consistently determined effective nucleon mass $M^*$ as it is the case in
relativistic mean-field models \cite{walecka}.   

\bigskip 

\bild{isorbitfig3.epsi}{14}
{\it Fig.\,6: Some diagrams representing the isoscalar nuclear spin-orbit 
interaction arising from irreducible two-pion exchange. The first two (last
two) diagrams are constructed from the $2\pi$-exchange triangle (crossed box)
graph of elastic  NN-scattering.}

\bigskip

In Fig.\,6 we show some diagrams which generate the isoscalar nuclear
spin-orbit strength related to irreducible two-pion exchange. By opening them
at the medium insertion (symbolized by the double line) ones see that the first
two Fock diagrams stem from the $2\pi$-exchange triangle graph of elastic
NN-scattering, whereas the last two Hartree and Fock diagrams are constructed  
from the $2\pi$-exchange crossed box graph of elastic NN-scattering. We have 
explicitly calculated the isoscalar nuclear spin-orbit strength $U_{ls}(p,k_f)$
generated by these diagrams using the techniques outlined in section 2. In an 
intermediate step the resulting isoscalar nuclear spin-orbit strength $U_{ls}
(p,k_f)$ can be written as an integral of the spin-orbit NN-amplitudes of the 
triangle and crossed box graph \cite{nnpap1} over a Fermi sphere of radius 
$k_f$. This expression can be further reduced to a one-dimensional integral of 
the form:
\begin{equation} U^{(2\pi)}_{ls}(p,k_f) = {1\over 16\pi^2 p^3} \int_{k_f-p}^{
k_f+p} dq\,q [k_f^2-(p+q)^2] [k_f^2-(p-q)^2] \Big\{ V_{SO}(q) +3 W_{SO}(q)
+2V_{SO}(0)\Big\}\,, \end{equation} 
with $V_{SO}(q)$ and $W_{SO}(q)$ the isoscalar and isovector spin-orbit
NN-amplitudes. The term $2V_{SO}(0)$ in eq.(15) belongs to Hartree type 
diagrams (with one closed and one open nucleon line) while the term $V_{SO}(q) 
+3 W_{SO}(q)$ summarizes the contributions from Fock type diagram (having just
one open nucleon line). The spin-orbit NN-amplitudes related to irreducible
$2\pi$-exchange have been calculated in section 4.2 of ref.\cite{nnpap1} using
dimensional regularization. If we use instead cut-off regularization the
linear divergences of the irreducible $2\pi$-exchange diagrams become visible. 
We find:  
\begin{equation} V_{SO}(q) = {3g_A^4 \over 64\pi M f_\pi^4} \bigg\{ -{11
\Lambda \over 6\pi} +m_\pi +{2m_\pi^2 +q^2 \over 2q} \arctan{q \over 2m_\pi}
\bigg\} \,, \end{equation} 
\begin{equation} W_{SO}(q) = {g_A^2 \over 64\pi M f_\pi^4} \bigg\{{(7g_A^2-8)
\Lambda \over 3\pi} +(1-g_A^2) \bigg[ m_\pi +{4m_\pi^2 +q^2 \over 2q} 
\arctan{q \over 2m_\pi} \bigg] \bigg\} \,, \end{equation}
which agrees up to the additive constants linear in the momentum cut-off
$\Lambda$ with the expressions in eqs.(22,23) of ref.\cite{nnpap1}. Note that
eqs.(16,17) include also the irreducible contributions of the $2\pi$-exchange
planar box graph (not shown in Fig.\,6). Inserting eqs.(16,17) into the "master
formula" eq.(15) one gets for isoscalar nuclear spin-orbit strength arising
from irreducible two-pion exchange:    
\begin{eqnarray} U^{(2\pi)}_{ls}(p,k_f) &=&{g_A^2\Lambda k_f^3 \over 3M(2\pi 
f_\pi)^4}\bigg(2+{19 g_A^2\over8}\bigg)+{g_A^2m_\pi^4\over35M(8\pi x)^3f_\pi^4}
\bigg\{ \bigg[ 24(7g_A^2 -4)\nonumber \\ &&  +28(5g_A^2 -4)(u^2+x^2) +35
\Big( {3g_A^2 \over 2}-2 \Big) (u^2-x^2)^2 \bigg]\ln{4+(u+x)^2\over 4+(u-x)^2} 
\nonumber \\ && +2(u+x)^3 \Big[(u^2-3u x+x^2)(28-14g_A^2+u^2+x^2)-10u^2x^2
\Big] \arctan{u+x \over 2} \nonumber \\ && +2(x-u)^3 \Big[(u^2+3u x+x^2)
(28-14g_A^2+u^2+x^2)-10u^2x^2 \Big] \arctan{u-x \over 2}\nonumber \\ && +2u x
\Big[ 12(4-7g_A^2)+(44-49g_A^2)(u^2+x^2)\nonumber \\ && -4\Big(u^4+x^4 +(22+105
g_A^2)u^2 x^2 \Big) \Big]\bigg\}\,.  \end{eqnarray}
At zero nucleon momentum $(p=0)$ this simplifies to:  
\begin{eqnarray}  U^{(2\pi)}_{ls}(0,k_f) &=&{g_A^2\Lambda k_f^3 \over 3M(2\pi
f_\pi)^4}\bigg( 2+{19 g_A^2 \over 8}\bigg) \nonumber \\ &&  +{g_A^2 k_f^2 \over
(4\pi)^3 M f_\pi^4} \bigg\{ -(3g_A^2+1) m_\pi k_f +\bigg[ (g_A^2-2) m_\pi^2
-{k_f^2 \over 2} \bigg] \arctan{k_f \over 2m_\pi} \bigg\} \,.\end{eqnarray}
Note the reciprocal dependence on the nucleon mass $M$ which indicates that
this nuclear spin-orbit strength is a relativistic effect. 

\bigskip

\bild{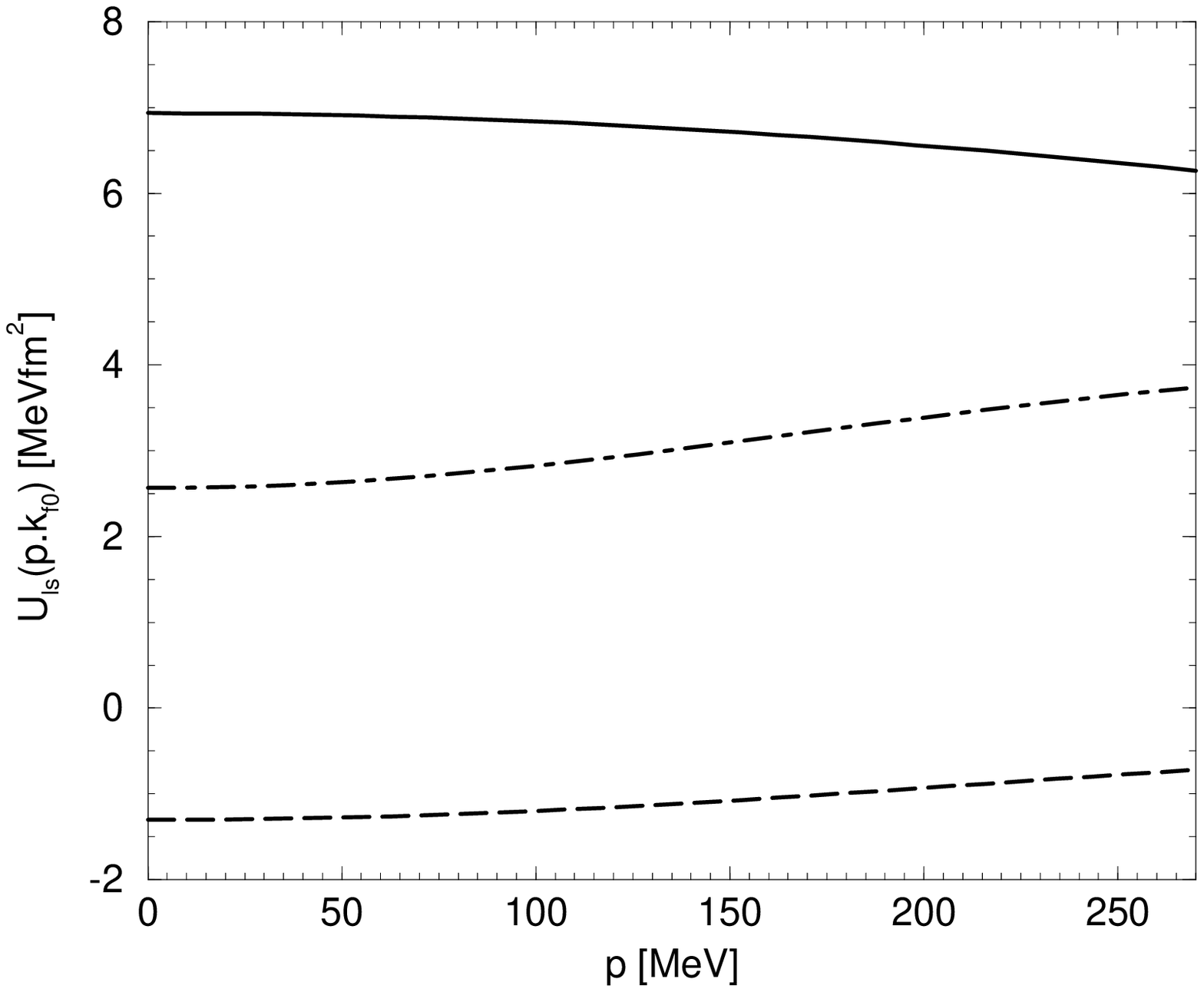}{13}
\vspace{-0.8cm}
{\it Fig.\,7: The full line shows the isoscalar nuclear spin-orbit strength 
$U^{(2\pi)}_{ls}(p,k_{f0})$ related to irreducible $2\pi$-exchange versus the
nucleon momentum $p$. The dashed line gives the combined result of the 
three-body diagrams in Fig.\,8 and the dashed-dotted line corresponds to the 
relativistic $1/M$-correction to the iterated $1\pi$-exchange Hartree diagram 
(a) in Fig.\,2.}
\bigskip

As in our previous work \cite{einpot} we  choose the value $\Lambda \simeq 
0.65\,$GeV for the momentum cutoff $\Lambda$. This gives at saturation density
$k_{f0}=272.7\,$MeV an isoscalar nuclear spin-orbit strength due to irreducible
$2\pi$-exchange of $U^{(2\pi)}_{ls}(0,k_{f0})= (16.20 -9.27)\,{\rm MeVfm}^2=
6.93\,{\rm MeVfm}^2$. This combination of $\Lambda$-dependent an finite terms
is less than 20\% of the total isoscalar spin-orbit strength coming from
iterated $1\pi$-exchange. Such a reduction factor is in accordance with the 
expectations from chiral power counting since the first/second term of $U^{(2
\pi)}_{ls}(0,k_f)$ in eq.(19) is suppressed by one/two powers of small momenta
in comparison to the results from iterated $1\pi$-exchange. The full line in 
Fig.\,7 shows the momentum dependence of the isoscalar nuclear spin-orbit 
strength  $U^{(2\pi)}_{ls}(p,k_{f0})$. One observes a very weak fall-off with
momentum from $p=0$ to $p=k_{f0}=272.7\,$MeV. 

\bigskip

\bild{isorbitfig4.epsi}{10}
{\it Fig.\,8: Three-body force Fock diagrams constructed from the (isovector)
Weinberg-Tomozawa $\pi\pi NN$-contact vertex. Analogous Hartree diagrams which
have an isospin factor zero in isospin-symmetric nuclear matter are not shown.}

\bigskip

By putting a second medium insertion to the intermediate nucleon propagator in
the first two diagrams in Fig.\,6 the first two graphs shown in Fig.\,8 result.
By opening them at the two medium insertions one sees that these graphs can be 
classified as three-body diagrams. With employment of one (isovector) 
Weinberg-Tomozawa $\pi\pi NN$-contact vertex one can construct a tree-level 
$2\pi$-exchange three-body diagram (NNN$\to$NNN). The three Fock diagrams in 
Fig.\,8 result then from this three-body diagram by closing two nucleon lines 
in all possible ways.\footnote{We discard the Hartree-type diagrams resulting
from this closing procedure since they have an isospin factor zero in 
isospin-symmetric nuclear matter.} Let us begin with the computation of the 
right Fock diagram in Fig.\,8 having a symmetrical arrangement of the two 
medium insertions. The $\pi\pi NN$-contact vertex (symbolized by the heavy dot)
contributes for the relevant kinematics first in form of a relativistic 
$1/M$-correction. After isolating the proportionality factor $i\,\vec \sigma 
\times \vec q$ from all spin-dependent terms one is left with an integral over
the product of two Fermi spheres of radius $k_f$. The latter can be solved in
closed form and one gets the following contribution from the third diagram in
Fig.\,8 to the isoscalar nuclear spin-orbit strength:
\begin{eqnarray}  U^{(3)}_{ls}(p,k_f) &=& {g_A^2 m_\pi^4 G(x,u)\over 24M (4\pi 
f_\pi x)^4} \bigg\{u\Big[ x^2(24+3x^2-8u^2)-3(1+u^2)^2 \Big]\nonumber \\ &&
+ 3(1+u^2-x^2)(1+2u^2+u^4 +10x^2-2u^2x^2+x^4) L(x,u) \nonumber
\\ && -24x^2 \Big[ \arctan(u+x)+  \arctan(u-x) \Big] \bigg\}\,, \end{eqnarray}
with the auxiliary function $G(x,u)$ defined after eq.(12). The computation of
the first two (topologically distinct) diagrams in Fig.\,8 is done in a similar
fashion. In this case one non-elementary integration is left over from the 
six-dimensional integral over the product of two Fermi spheres of radius
$k_f$. We find the following expression for the isoscalar nuclear spin-orbit 
strength:
\begin{eqnarray}U^{(3)}_{ls}(p,k_f) &=&{g_A^2m_\pi^4\over M(4\pi f_\pi)^4 x^3} 
\int_0^u d\xi \bigg\{4\xi^2 \Big[ \arctan(u+\xi)+  \arctan(u-\xi) \Big]
\nonumber \\ &&  \times  \Big[ (1+\xi^2-x^2)L(\xi,x)-x\Big]  + {u x\over 24}
\Big[ 9\xi^{-2} (1+u^2)^2(1+x^2) -87 \xi^4 \nonumber \\ &&-\xi^2(93+146u^2+
99x^2) -3(31+40u^2+u^4+30x^2+38u^2x^2)\Big]\nonumber \\ &&+ {1\over 8} L(\xi,x)
L(\xi, u) \Big[ \xi^6 (62x^2+57u^2-89)-29 \xi^8 + 3\xi^{-2}(1+u^2)^3(1+x^2)^2
\nonumber \\ &&+ 3 \xi^4(12u^2-62-9u^4 +8x^2-42 u^2x^2 -11x^4)  \nonumber \\ 
&&+ \xi^2 (69u^2 x^4 -63x^4+66u^4x^2 +294 u^2x^2 -36 x^2 -u^6 -69 u^4 -90 u^2
-154 ) \nonumber \\ &&-(1+u^2) (25+35u^2-2u^4 +56x^2 +82 u^2x^2 +2u^4 x^2 +27
x^4 +39 u^2 x^4) \Big]  \nonumber \\ &&+ {x\over 8} L(\xi,u) \Big[(1+u^2)(28+
41u^2+u^4 +27x^2 +39u^2 x^2) -3\xi^{-2}(1+u^2)^3(1+x^2) \nonumber \\ &&+29\xi^6
+3\xi^4 (20-19u^2+11x^2) + 3\xi^2(42+7u^2 +9u^4 +21x^2
-23u^2x^2)\Big] \nonumber \\ &&+ {u\over 24} L(\xi,x) \Big[
87\xi^6 +2\xi^4(90+73 u^2-93 x^2) -9\xi^{-2}(1+u^2)^2(1+x^2)^2  \nonumber \\
&&+  \xi^2(186+266u^2+3u^4 +114 x^2 -260 u^2x^2+99x^4 )  \nonumber \\ &&+6(
14+17u^2 -u^4+31x^2 +40u^2 x^2 +u^4 x^2 +15x^4 +19 u^2 x^4) \Big] \bigg\}\,, 
\end{eqnarray}
to which the first and second diagram in Fig.\,8 have contributed equally.
Numerical evaluation of eqs.(20,21) gives at nuclear matter saturation density
$k_{f0}=272.7\,$MeV and zero nucleon momentum $U^{(3)}_{ls}(0,k_{f0})=(-0.20-
1.10)\,{\rm MeVfm}^2=-1.30\,{\rm MeVfm}^2$. This is a very small effect 
compared to iterated $1\pi$-exchange. The momentum dependence of the combined
result from eqs.(20,21) for $U^{(3)}_{ls}(p,k_{f0})$ is shown by the dashed 
line in Fig.\,7. One observes a weak increase with momentum from $p=0$ to 
$p=k_{f0}$. In the heavy baryon framework the relativistic $1/M$-corrections 
calculated here are generated by (nucleon-momentum dependent) vertices from the
second order chiral Lagrangian ${\cal L}^{(2)}_{\pi N}$ (proportional to
$1/M$). The other $\pi\pi NN$-contact vertices proportional to the low-energy 
constants $c_{1,2,3,4}$ give rise to spin-orbit terms first one order higher in
the small momentum expansion. An order of magnitude estimate of their effect on
$U_{ls}(p,k_f)$ can be obtained by inserting the spin-orbit NN-amplitudes
$W_{SO}(q)$ and $V_{SO}(q)$ in eqs.(13,15,16) of ref.\cite{2pi2loop} (which do 
not include regularization dependent polynomial pieces) into the "master 
formula" eq.(15).  

There are of course many more contributions to the isoscalar nuclear spin-orbit
strength at fourth order in small momenta (having the common prefactor $m_\pi^4
/M$). In order to get an idea about the typical size of other such effects we
consider now the first relativistic $1/M$-corrections to the dominant iterated
$1\pi$-exchange Hartree diagrams in Fig.2. In order to obtain conveniently all
$1/M$-corrections from these diagrams one starts with fully relativistic
interaction vertices and fully relativistic pion- and nucleon propagators 
sandwiched between Dirac spinors. Next one performs an expansion in powers of 
$1/M$ to the appropriate order and solves the remaining (loop)-integrals. In 
the case of diagram (a) in Fig.\,2 with one medium insertion one ends up with 
the following $1/M$-correction to the isoscalar nuclear spin-orbit strength:  
\begin{eqnarray}  U_{ls}^{(a)}(p,k_f)|_{M^{-1}} &=& {g_A^4 m_\pi^4 \over M(8\pi
x)^3f_\pi^4} \bigg\{ u x\bigg( 9u^4+9x^4 -6u^2x^2 +{27+43 u^2 +483 x^2 \over
20} \bigg) \nonumber \\ && -{2\over 5} (28u^5+45x^3 +10u^2 x^3 +18 x^5)
\arctan(u+x)  \nonumber \\ && +{2\over 5} (28u^5-45x^3 -10u^2 x^3 -18 x^5)
\arctan(u-x)\nonumber \\ && + \bigg[ {7\over 2}(3x^2-u^2) -{27 \over 20} +{15
\over 4} (3u^4+2u^2x^2-5x^4) \nonumber \\ && -9(u^2-x^2)^2(u^2+x^2) \bigg] x
L(x,u) \bigg\} +{g_A^4 \Lambda k_f^3 \over 4M(2\pi f_\pi)^4} \,.\end{eqnarray}
In this expression we have subtracted the irreducible part $g_A^4 k_f^3 (6
\Lambda -5\pi m_\pi)/(128 \pi^4 M f_\pi^4)$ which is already accounted for in 
eqs.(15,18) by the spin-orbit NN-amplitude $V_{SO}(0)$ derived from the
planar $2\pi$-exchange box graph. Inserting the cut-off scale of $\Lambda \simeq
0.65 \,$GeV one gets $U_{ls}^{(a)}(0,k_{f0})|_{M^{-1}} =(-0.85+3.41)\,{\rm
MeVfm}^2  = 2.56\,{\rm  MeVfm}^2$ which is a small $-5\%$ correction to the 
leading order contribution $U_{ls}^{(a)}(0,k_{f0})|_{M}=-52.96\,{\rm MeVfm}^2$ 
\cite{spinbahn}. The momentum dependence of $U^{(a)}_{ls}(p,k_{f0})|_{M^{-1}}$ 
is shown by the dashed-dotted line in Fig.\,7. One observes a moderate increase
with momentum from $p=0$ to $p=k_{f0}$.  

In order to avoid very lengthy analytical expressions, we present here only the
first relativistic $1/M$-corrections to the Hartree diagrams (b) and (c) in
Fig.\,2 evaluated at zero nucleon momentum ($p=0)$. In the case of diagram (b)
one finds: 
\begin{equation} U_{ls}^{(b)}(0,k_f)|_{M^{-1}} = {g_A^4 m_\pi^4 u^2 \over M(2
\pi f_\pi )^4} \bigg\{ \bigg( {15 \over 8u} -u \bigg) \arctan u +{ 4u^6 +16 u^4
-51 u^2 -45 \over 24(1+u^2)^2} \bigg\} \,. \end{equation} 
The resulting numerical value $U_{ls}^{(b)}(0,k_{f0})|_{M^{-1}}= -1.06 \,{\rm 
MeVfm}^2$ is again a small $-3\%$ correction to the leading term $U_{ls}^{(b)}
(0,k_{f0})|_{M} = 37.35\,{\rm MeVfm}^2$ \cite{spinbahn}. The analogous result 
for the Hartree diagram (c) reads: 
\begin{eqnarray}  U_{ls}^{(c)}(0,k_f)|_{M^{-1}} &=& {g_A^4 m_\pi^4 \over M(4\pi
f_\pi)^4} \bigg\{ (6+4u^2) \ln(1+u^2) +8u(2u^2-5) \arctan u  \nonumber \\ &&
+{u^2(2-u^2)\over 3(1+u^2)^3} (51+133u^2 +103 u^4 +13u^6) \nonumber \\ &&+{2
\over u}\int_0^u d\xi \,{\xi^4(u-\xi) \over (1+\xi^2)^4} \ln{u+\xi\over u-\xi} 
\Big[u(2-u^2) \xi^4 -5\xi^7 -20 \xi^5 \nonumber \\ && -5u^3 +(4u^2-27) \xi^3 
-2u(3u^2+5) \xi^2 +4u^2 \xi \Big] \bigg\} \,,\end{eqnarray}
which leads to an even smaller numerical value of $U_{ls}^{(c)}(0,k_{f0})|_{
M^{-1 }}= 0.10 \,{\rm MeVfm}^2$. 

In summary, we have calculated here several relativistic $1/M$-corrections to
the isoscalar nuclear spin-orbit strength $U_{ls}(p,k_f)$ generated by chiral
$2\pi$-exchange. As expected from chiral power counting these higher order
terms are significantly smaller than the unique leading contribution from
iterated $1\pi$-exchange \cite{spinbahn}. Its role in finite nuclei should be
clarified by implementing it into nuclear structure calculations.

\end{document}